\documentclass[a4paper,twoside]{article}

\usepackage{tikz}
\usetikzlibrary{arrows.meta, positioning, shapes.geometric, calc}

\usepackage{epsfig}
\usepackage{subcaption}
\usepackage{calc}
\usepackage{amssymb}
\usepackage{amstext}
\usepackage{amsmath}
\usepackage{amsthm}
\usepackage{multicol}
\usepackage{pslatex}
\usepackage{algorithm2e}
\usepackage[bottom]{footmisc}
\usepackage{url}
\usepackage{SCITEPRESS}     % Please add other packages that you may need BEFORE the SCITEPRESS.sty package.
\usepackage{cite}
\usepackage{makecell}
\usepackage{booktabs} % for nicer tables
\usepackage{multirow} % optional, for multirow cells

\begin{document}

\title{Detecting and Explaining Malware Family Evolution Using Rule-Based Drift Analysis}

\author{\authorname{Olha Jurečková\sup{1}\orcidAuthor{0000-0002-8858-4826},  and Martin Jureček\sup{1}\orcidAuthor{0000-0002-6546-8953}}
\affiliation{\sup{1}Faculty of Information Technology, Czech Technical University in Prague, \mbox{Thákurova 9}, Prague, 16000, Czech Republic}
\email{\{jurecolh, martin.jurecek\}@fit.cvut.cz}}


\abstract{Malware detection and classification into families are critical tasks in cybersecurity, complicated by the continual evolution of malware to evade detection. This evolution introduces concept drift, in which the statistical properties of malware features change over time, reducing the effectiveness of static machine learning models. Understanding and explaining this drift is essential for maintaining robust and trustworthy malware detectors. In this paper, we propose an interpretable approach to concept drift detection. Our method uses a rule-based classifier to generate human-readable descriptions of both original and evolved malware samples belonging to the same malware family. By comparing the resulting rule sets using a similarity function, we can detect and quantify concept drift. Crucially, this comparison also identifies the specific features and feature values that have changed, providing clear explanations of how malware has evolved to bypass detection. Experimental results demonstrate that the proposed method not only accurately detects drift but also provides actionable insights into the behavior of evolving malware families, supporting both detection and threat analysis.}

\onecolumn \maketitle \normalsize \setcounter{footnote}{0} \vfill

\section{\uppercase{Introduction}}
\label{sec:introduction}

Malware detection and classification into families are fundamental challenges in cybersecurity, as the continuous evolution of malicious code to evade detection significantly hampers accurate classification, thereby undermining defense mechanisms and delaying timely incident response \cite{bensaoud2024survey, eren2024catch}. The classification of malware into families is fundamental to understanding the behavior, origin, and evolution of malicious code, as it enables more accurate identification of variants and enhances the effectiveness of detection systems and response strategies in the face of continuously evolving threats \cite{rieck2008learning, hassen2017malware}. Modern malware rarely exists in isolation, but rather as part of an evolving family that continuously mutates to evade detection systems, posing significant challenges to traditional approaches and motivating the use of advanced machine learning techniques and large-scale empirical analyses to uncover behavioral patterns, inheritance relationships, and evolutionary dynamics essential for accurate classification and effective threat mitigation \cite{gupta2009empirical, wadkar2020detecting, tupadha2022machine}.

Malware is constantly evolving, and this evolution leads to concept drift, i.e., changes in the statistical properties of malware over time, which significantly challenges traditional detection systems. To remain effective, machine learning–based detectors must adapt both their models and the features they use \cite{ceschin2023fast}. In malware detection, concept drift occurs when the characteristics that define malicious behavior shift due to changes introduced by malware authors, such as packing techniques, control-flow obfuscation, or altered API usage patterns. Detecting and understanding this drift is essential for maintaining the effectiveness of machine learning–based malware detectors.

Beyond detection accuracy, interpretability is increasingly recognized as a key requirement in malware detection systems. While black-box models may achieve high predictive performance, they offer little insight into why a sample is classified as malicious, limiting analyst trust and actionable understanding. Interpretable models are especially important in adversarial domains such as malware analysis, where understanding the rationale behind a classification can guide remediation efforts and enhance model robustness.

This paper proposes a novel approach for detecting and explaining concept drift in malware families over time. To enable both detection and interpretability, we leverage a rule-based classifier, which produces a human-readable set of logical rules describing malware samples. Applying the same rule-based classifier to evolved samples from the same malware family allows to compare the resulting rule sets using a similarity function, and detect and quantify the concept drift. Since the rule-based classifier produces explicit formulas, our drift detector inherently supports explanation: it identifies specific conditions or features that differ between original and evolved samples, shedding light on the malware’s evolutionary tactics. The proposed framework thus provides a dual benefit: it captures concept drift with quantitative rigor, and it does so in an interpretable way that supports security analysis and incident response.

The main contributions of this paper are summarized as follows:
\begin{itemize}
    \item We propose a novel framework for detecting concept drift in malware families that explicitly emphasizes interpretability by leveraging a rule-based classifier.
    \item We introduce a rule-set comparison mechanism based on similarity measures to detect and quantify evolutionary changes within malware families over time.
    \item We demonstrate that concept drift detection can be inherently explainable when expressed through human-readable logical rules, enabling direct identification of features and conditions responsible for malware evolution.
    \item We experimentally validate the proposed approach on six malware families, achieving an overall concept drift detection accuracy of 92.08\%.
\end{itemize}

This paper is organized as follows. Section \ref{sec:rl} reviews related work on malware evolution and concept drift detection. Section \ref{sec:back} provides the necessary background, including the malware families used in our experiments, the rule-based classifier, and the adversarial malware generator MAB-malware. Section \ref{sec:proposed} presents our proposed approach for detecting and explaining malware family evolution. Section \ref{sec:experiments} describes the experimental setup and results. Finally, Section \ref{sec:conclusion} concludes the paper by summarizing the main findings on rule-based concept drift detection in malware families and outlining directions for future research.

\section{Related Work}
\label{sec:rl}

Concept drift, a shift in the data distribution or decision boundary, has received attention in adversarial domains where attackers adapt to evade detection~\cite{gama2014survey}. In the context of malware, drift can manifest through obfuscation, packing, or structural code changes~\cite{mohaisen2013avmeter}. Tools such as \textit{Transcend}~\cite{jordaney2017transcend} and \textit{ADClust}~\cite{kwon2020adclust} have been proposed for detecting drift in classifiers, but they typically use black-box models and do not provide insight into \textit{how} or \textit{why} drift occurs.

A few recent works have attempted to interpret the effects of drift. Jordaney et al.~\cite{jordaney2017transcend} introduced a framework for detecting drift in malware classifiers using ensemble disagreement, while Woodbridge et al.~\cite{woodbridge2016predicting} proposed techniques for identifying semantic drift through behavior-based clustering. However, these approaches either require labeled data streams or fail to explain the drift at the feature level.

In contrast, our work emphasizes interpretable concept drift detection using a rule-based classifier, enabling both quantification and explanation of how malware features evolve over time. By comparing the logical structure of rule sets generated at different time points, we detect not only that drift has occurred, but also which features and conditions are responsible, an aspect that has been largely overlooked in existing literature.

In \cite{singh2012tracking} the authors investigated concept drift in malware detection by proposing tracking methods and analyzing different forms of malware evolution. The study aimed to detect drift by comparing classifier performance on original and newly timestamped datasets. Two drift-monitoring measures were introduced, relative temporal similarity and metafeatures, applied to static features using cosine similarity. The authors used instruction mnemonics and byte 2-grams to assess code evolution across three real-world malware families. Their findings indicated negligible drift in mnemonic 2-grams.

The authors of \cite{jordaney2017transcend} presented Transcend, a completely parametric and adjustable statistical framework for concept drift detection. It is based on the Conformal Evaluator (CE), which uses non-conformity metrics from the machine learning algorithm being evaluated to statistically evaluate the quality of predictions. Transcend formulates drift detection as a parametric process along two dimensions: the desired performance level and the proportion of samples an analyst is willing to manually inspect per epoch.  These parameters serve as degrees of freedom to guide drift detection. Thresholds separating correct from incorrect classifications are derived from quality metrics computed during training and are applied during deployment, even in the absence of labeled data.

The authors of \cite{zola2023temporal} conducted a temporal analysis of distribution shifts in malware classification and proposed a three-step forensic approach to investigate model failures due to concept drift. First, they examined concept drift using a rolling window strategy for training data selection. Second, they evaluated model drift based on the amount  of temporal information used in the training dataset. Third, they performed a detailed misclassification and feature analysis to interpret drift-related failures. Their evaluation was performed on multi-class classifiers that utilize structural embeddings extracted from malware Control Flow Graphs (CFGs), employing three distinct classifier families, each implemented in two configurations.

In \cite{augello2025hybrid}, the authors propose a two-level malware detection framework combining lightweight on-device analysis with selective cloud-based analysis. A self-evaluation agent detects potential misclassifications due to concept drift and triggers remote analysis when needed. Ensembles of random forests are used, with the KNORA-U algorithm dynamically selecting classifiers for majority-vote predictions. Experiments show that this approach effectively mitigates concept drift while maintaining efficient detection.

In contrast to the predominantly statistical and learning-based approaches discussed above, our work focuses on concept drift detection using a rule-based classifier, thereby explicitly prioritizing interpretability and transparency. While existing methods rely on complex neural architectures, ensemble models, or latent-space clustering, often operating as black boxes, our approach enables direct inspection of decision rules and their evolution over time. This facilitates a clearer understanding of how and why concept drift is detected, which is particularly valuable in security-critical settings where analyst trust and explainability are essential. To the best of our knowledge, this work represents one of the first studies of concept drift detection in malware analysis using an interpretable, rule-based framework, bridging the gap between effective drift detection and practical forensic analysis.

\section{Background}
\label{sec:back}

This section introduces the malware families used for concept drift detection. It further provides background information on rule-based classification, which is employed to describe a given malware family using a set of rules. Finally, it presents the MAB-malware adversarial generator, which is used to create new iterations of malware families containing samples that successfully evade the given target classifier.

\subsection{Malware families}
\label{sec:fam}

The concept of a malware family is commonly used to group malicious programs that share significant structural, behavioral, or code-level similarities. A malware family can be defined as a set of malware variants originating from a common code base or builder, often characterized by shared functionality, propagation mechanisms, or distinctive code patterns. Membership in a family reflects the fact that malware is rarely created in isolation; attackers typically reuse and adapt existing code to produce new variants. This results in clusters of related samples that retain core malicious capabilities while differing in obfuscation techniques, payload delivery, or evasion strategies.

In the following, we provide a detailed description of six malware families used in our experiments. Additional information about these malware families can be found in \cite{kaspersky_threats}.

\begin{itemize}
\item Agensla is a Trojan-PSW program designed to steal user account information, such as logins and passwords, from infected computers, specifically targeting the Microsoft .NET Framework platform (MSIL).

\item DCRat is a modular backdoor malware belonging to the Dark Crystal RAT family (DCRat for short) and is classified as Backdoor.MSIL.DCRat. Beyond its core backdoor functionality, it can load additional modules to extend its operational capabilities. The malware is commonly distributed using deceptive tactics, such as fraudulent or compromised YouTube accounts that advertise gaming cheats or cracks, with download links leading victims to malicious payloads.

\item Makoob is classified as a Trojan spyware program that covertly monitors user activity, including active processes, screenshots, and keystrokes. The captured data is then transmitted to the attacker via various network channels, such as email, FTP, and HTTP.

\item Mokes, also known as Smoke Loader, is a modular Win32 backdoor distributed via the Cutwail spam botnet that primarily functions as a loader for additional malicious payloads, such as Trojan-Ransom.Win32.Cryptodef. Its modular architecture enables dynamic extension of capabilities, including hosts-file modification, credential theft, interception of browser-input data, and execution of arbitrary shellcode on infected systems. 

\item Strab is a Windows Trojan that records keystrokes, captures screenshots, and enumerates active processes to collect sensitive information from files and the system registry. The collected data is typically transmitted to a remote attacker via email, FTP, or HTTP requests. 

\item Taskun is classified as a Trojan-Downloader that facilitates the installation of additional malicious software, including updated variants of Trojans and adware, on compromised systems. Once retrieved from remote servers, the downloaded components are either executed immediately or configured to run automatically at system startup.

\end{itemize}

While malware families consist of numerous individual samples, the descriptions in this work summarize representative behavioral characteristics that are consistently observed across variants of each family. These summaries are not intended to describe a single malware instance, but rather to provide contextual background on the typical functionality and threat model associated with each malware family considered in the experimental evaluation.

\subsection{Rule-Based Classification}
\label{pravidla}

Interpretability is a critical requirement in malware detection and family classification. While black-box models, such as deep neural networks, can achieve high accuracy, their inner workings are often opaque, making it difficult for malware researchers to understand why a sample is assigned to a particular family. Transparent decision logic is essential in this domain, as security analysts need to validate detection results, attribute attacks to known families, and derive actionable intelligence from classification outcomes. Rule-based classifiers address this need by providing human-readable conditions that explicitly capture the distinguishing characteristics of malware families.

A \emph{condition} $c$ is formally defined as
\begin{equation}
  c \equiv (x \;\odot\; h),
\end{equation}
where $x$ denotes a feature of a malware sample, $\odot$ is a relational operator 
(e.g., $=, \neq, >, <, \geq, \leq$), and $h$ is the target value associated with 
the feature $x$. 
A \emph{rule} $r$ is defined as a conjunction of $m$ conditions:
\begin{equation} \label{rule}
  r \equiv c_1 \wedge c_2 \wedge \dots \wedge c_m,
\end{equation}
where $c_1, \dots, c_m$ are individual conditions and $m$ denotes the number of conditions in the rule $r$.  
The \emph{size} of the rule is defined as $m$, i.e., the total number of constituent conditions.

One of the most influential algorithms for learning rule sets is RIPPER (Repeated Incremental Pruning to Produce Error Reduction). RIPPER \cite{cohen1995} is an inductive rule learner that incrementally constructs rules to separate malware from benign samples or to discriminate between different malware families. During training, it generates candidate rules, prunes irrelevant conditions, and repeatedly optimizes the rule set to minimize errors on unseen data. The output is a compact, interpretable collection of if–then statements that directly explain classification decisions. This property makes RIPPER particularly well suited for malware research, where understanding and explaining why a sample is associated with a given family is often as important as the classification itself.

In our work, we used the \texttt{wittgenstein} library\footnote{\url{https://github.com/imoscovitz/wittgenstein}}, which provides an implementation of RIPPER. The output of this algorithm is a rule set defined as a disjunction of rules, which are described in \eqref{rule}. In the following text, we will simplify the terminology and refer to a rule set simply as rules. Below is an example of rules describing the Agensla family, where each sample is represented using only three features: $f_1, f_2$, and $f_3$.\\

\begin{flushleft}
\small
\texttt{[[$f_3$=72 AND $f_1$ in [10,48] AND $f_2$=1] OR}\\
\texttt{[$f_3$=72 AND $f_1\geq$48]} \texttt{OR [$f_3$=72 AND $f_1$ in [10,48] OR [$f_3$=72 AND $f_1 \leq 6$ AND $f_2$=0]} \texttt{OR [$f_2$=1 AND $f_1 \leq 6$ AND $f_3$=72]} \texttt{OR [$f_3$=72 AND $f_1$ in [7,8] AND $f_2$=1]]}
\end{flushleft}

\subsection{MAB-malware}

MAB-Malware \cite{song2021mabmalware} is a reinforcement learning-based adversarial malware generator that employs a multi-armed bandit (MAB) agent to identify minimal sets of binary modifications that cause a target classifier to mislabel malicious samples as benign. The method proceeds in two phases: 
\begin{enumerate}
\item[(1)] an attack phase in which the MAB agent iteratively applies candidate macro- and micro-manipulations until evasion is achieved or a budget is exhausted, and
\item[(2)] a minimization phase in which each applied modification is re-tested and removed if it is not required for successful evasion.
\end{enumerate} 
Since the MAB formulation treats actions as independent (no ordering or dependency is assumed), post-hoc pruning effectively reduces perturbation while preserving evasion.

Typical manipulations include appending benign data (overlay/section), adding or renaming sections, zeroing certificate or debug fields, corrupting optional header checksums, and semantically preserving code transformations. We used MAB-Malware to generate adversarial malicious examples against an EMBER-based classifier \cite{anderson2018ember}.

\section{Proposed Approach: Detecting Concept Drift in Malware Families Using Rule-Based Classifiers} \label{sec:proposed}

In this work, we propose a method to detect concept drift in malware families by leveraging 
rule-based classifiers. Concept drift occurs when the statistical properties of a malware family 
change over time, for instance due to modifications introduced by malware authors or automated 
adversarial generation techniques. Detecting such drift is crucial for maintaining effective 
detection systems and for understanding the evolution of malware.

Our approach proceeds in three main stages. First, we generate a set of rules describing 
the original malware family. These rules are learned using a rule-based classifier, such as 
RIPPER, from features extracted from a representative set of samples. Each rule $r$ is defined 
as a conjunction of conditions $c_1 \wedge \dots \wedge c_m$, capturing the structural and 
behavioral properties that characterize the family.

Next, we evolve the malware family by generating adversarial variants using an automated 
malware generator. The generated samples are intended to preserve the malicious functionality 
while introducing changes that may affect the classifier’s decision boundaries. From this 
evolved set of samples, we generate a second set of rules using the same rule-based learning 
process. These rules describe the updated characteristics of the family after evolution.

Finally, we compare the rules obtained for the original and evolved families using a 
\emph{rule distance function}, which quantifies the differences between sets of rules. Significant differences in 
the rules indicate that the family has undergone concept drift. Formally, if $R_{\text{orig}}$ 
and $R_{\text{adv}}$ denote the rule sets for the original and adversarial families, respectively, 
the drift score can be expressed as
\begin{equation}
     f_{\text{distance}}(R_{\text{orig}}, R_{\text{adv}}),
\end{equation}
where $f_{\text{distance}}$ is a suitable distance metric capturing rule dissimilarity. Using this distance metric, we can calculate the degree of dissimilarity between sets of rules, which allows us to quantify concept drift.

Specifically, the comparison between two RIPPER rules is performed using the normalized Hamming distance

\begin{equation}
\label{eq:rule_distance}
d(x, y) = \frac{1}{n} \sum_{i=1}^{n} \mathbf{1}_{\{x_i \neq y_i\}},
\end{equation}

where

\begin{itemize}
    \item $x, y \in \{0,1\}^n$ are two binary vectors,
    \item $n$ is the length of the vectors, where the component $x_i = 1$ (resp.\ $y_i = 1$) indicates that the $i$-th sample is detected by the rule, and $0$ otherwise,
    \item $\mathbf{1}_{\{x_i \neq y_i\}}$ is an indicator function that equals $1$ if $x_i \neq y_i$ and $0$ otherwise.
\end{itemize}

This metric quantifies the proportion of positions at which the corresponding binary vectors differ, reflecting the frequency with which the rule sets make different decisions. It is naturally bounded between 0 and 1 and provides a straightforward, interpretable measure of disagreement between rule sets, making it well-suited for analyzing evolutionary drift in malware families.

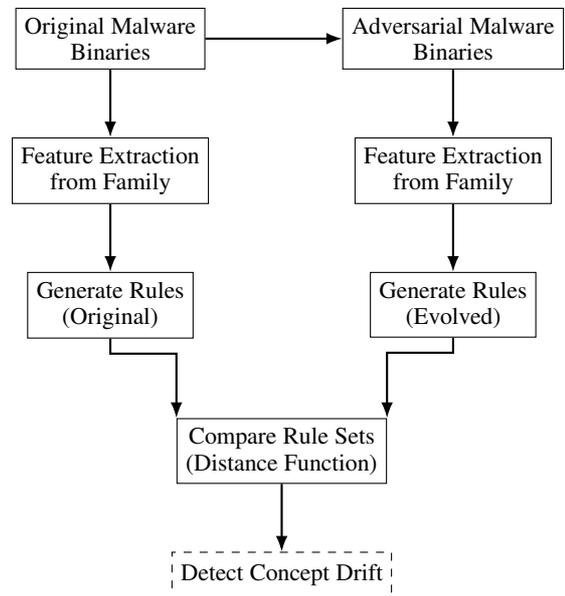
\begin{figure}[htbp]
\begin{center} % or \centering
\begin{tikzpicture}[
    node distance=0.9cm and 1.8cm,
    process/.style={rectangle, draw, minimum width=2.1cm, minimum height=0.6cm, align=center, font=\small},
    data/.style={rectangle, draw, dashed, minimum width=2.1cm, minimum height=0.6cm, align=center, font=\small},
    arrow/.style={-{Latex[scale=0.9]}, thick}
]

% Sources (top row)
\node[process] (binaries) {Original Malware \\ Binaries};
\node[process, right=of binaries] (adv_gen) {Adversarial Malware \\ Binaries};

% Feature extraction row
\node[process, below=of binaries] (extract_orig) {Feature Extraction \\ from Family};
\node[process, below=of adv_gen] (extract_adv) {Feature Extraction \\ from Family};

% Rules row
\node[process, below=of extract_orig] (rules_orig) {Generate Rules \\ (Original)};
\node[process, below=of extract_adv] (rules_adv) {Generate Rules \\ (Evolved)};

% Comparison & drift (centered under both rule nodes)
\node[process, below=1.5cm of $(rules_orig)!0.5!(rules_adv)$] (compare) {Compare Rule Sets \\ (Distance Function)};
\node[data, below=of compare] (drift) {Detect Concept Drift};

% Arrows: original branch
\draw[arrow] (binaries) -- (extract_orig);
\draw[arrow] (extract_orig) -- (rules_orig);

% Arrow: original binaries -> adversarial generator
\draw[arrow] (binaries.east) -- (adv_gen.west);

% Arrows: adversarial branch
\draw[arrow] (adv_gen) -- (extract_adv);
\draw[arrow] (extract_adv) -- (rules_adv);

% Arrows: merging branches into compare
\draw[arrow] (rules_orig.south) -- ++(0,-0.2) -| (compare.north west);
\draw[arrow] (rules_adv.south) -- ++(0,-0.2) -| (compare.north east);

% Final arrow
\draw[arrow] (compare) -- (drift);

\end{tikzpicture}
\end{center}
\caption{Overview of the proposed concept drift detection framework}
  \label{fig:drift-framework}
\end{figure}

This methodology allows us to systematically detect and quantify changes in malware families, 
providing insights into their evolution and enabling timely updates of detection models. 
By relying on interpretable rules, the approach also offers explanations for observed drift, 
supporting forensic analysis and threat attribution. Figure~\ref{fig:drift-framework} illustrates the procedure for detecting concept drift based on the distance between sets of rules.\\

\section{\uppercase{Experiments}}
\label{sec:experiments}

This section presents the experimental setup, describes the experiments, and reports the results of concept drift detection.

\subsection{Experimental Setup}
%In \cite{balik2025rawmal} the authors introduced the dataset that includes labels for both malware families and malware types. The RawMal-TF dataset consists of over 250,000 malware samples across 17 families and 14 malware categories, amounting to more than 160 GB of Windows PE malware files. To create the dataset, the authors collected raw binary samples from sources such as VirusShare, VX Underground, and MalwareBazaar. Family labels were derived from the binaries’ filenames, while type-level annotations were added using data from ClarAVy.

In this work, we used binary files from the RawMal-TF dataset \cite{balik2025rawmal}, which contains malware samples categorized into families and by type (e.g., virus, worm). Using the MAB-malware adversarial generator, we produced adversarial modifications of these malware samples and retained only those that successfully evaded the EMBER classifier. Since RIPPER also requires benign samples for training, their feature vectors were obtained from the EMBER dataset \cite{2018arXiv180404637A}. In the experimental part, we worked with six malware families, Agensla, DCRat, Makoob, Mokes, Strab, and Taskun, which are described in Section~\ref{sec:fam}. Table~\ref{tab:sample_overview} summarizes the numbers of malware samples used in our experiments.

\begin{table}[htbp]
\centering
\caption{Overview of malware families and sample counts}
\label{tab:sample_overview}
\begin{tabular}{l|c|c}
\hline
Family & \makecell{Number of\\original samples} & \makecell{Number of\\adversarial samples} \\
\hline
Agensla  & 8,418 & 2,558 \\
DCRat     & 1,026 &  1,010\\
Makoob  & 2,414 & 626 \\
Mokes  & 2,216 & 2058 \\
Strab  & 2,191 &  1,596\\
Taskun  & 4,888 &  1,015\\
\hline
\end{tabular}
\end{table}

The feature set used in this work is based on the LIEF library\footnote{\url{https://github.com/lief-project/LIEF}}, a cross-platform library for parsing and modifying executable formats such as PE, ELF, and Mach-O, with bindings for multiple programming languages. In our work, we employ this library to obtain a static, fixed-length representation of Windows Portable Executable (PE) files by combining metadata, header information, section statistics, imported and exported functions, and byte-level distributions. The resulting representation is designed to efficiently capture both structural and content-based characteristics of binaries, enabling large-scale machine learning-based malware detection.

The objective of this work is to experimentally verify that the proposed approach is capable of detecting concept drift based on distances between rule sets. The experimental procedure is outlined as follows.\\

\begin{enumerate}
    \item From the RawMal-TF dataset, we extracted the six aforementioned malware families (hereafter referred to as the \emph{original families}).
    \item Each original family was processed using the MAB-malware adversarial generator, which modifies malware samples to make them more difficult for the EMBER classifier to detect (hereafter referred to as the \emph{adversarial families}).
    \item Each original family was randomly divided into two equally sized subsets, denoted as \textit{set1} and \textit{set2}. Rule sets were then computed separately for \textit{set1} and \textit{set2} using the RIPPER rule-based classifier.
    \item The distance between the rule sets obtained in the previous step was computed according to Equation \eqref{eq:rule_distance}.
    \item Rule sets were computed from the adversarial samples using the RIPPER rule-based classifier.
    \item The distance between the rule sets obtained in Step~5 and those obtained in Step~3 was computed according to Equation \eqref{eq:rule_distance}.
    \item Concept drift detection was performed based on differences in the rule-set distances obtained in Step~6.
\end{enumerate}

\subsection{Experimental Results}

\begin{figure*}[htbp]
    \centering
    % First row
    \begin{subfigure}[b]{0.47\textwidth}
        \centering
        \includegraphics[width=\textwidth]{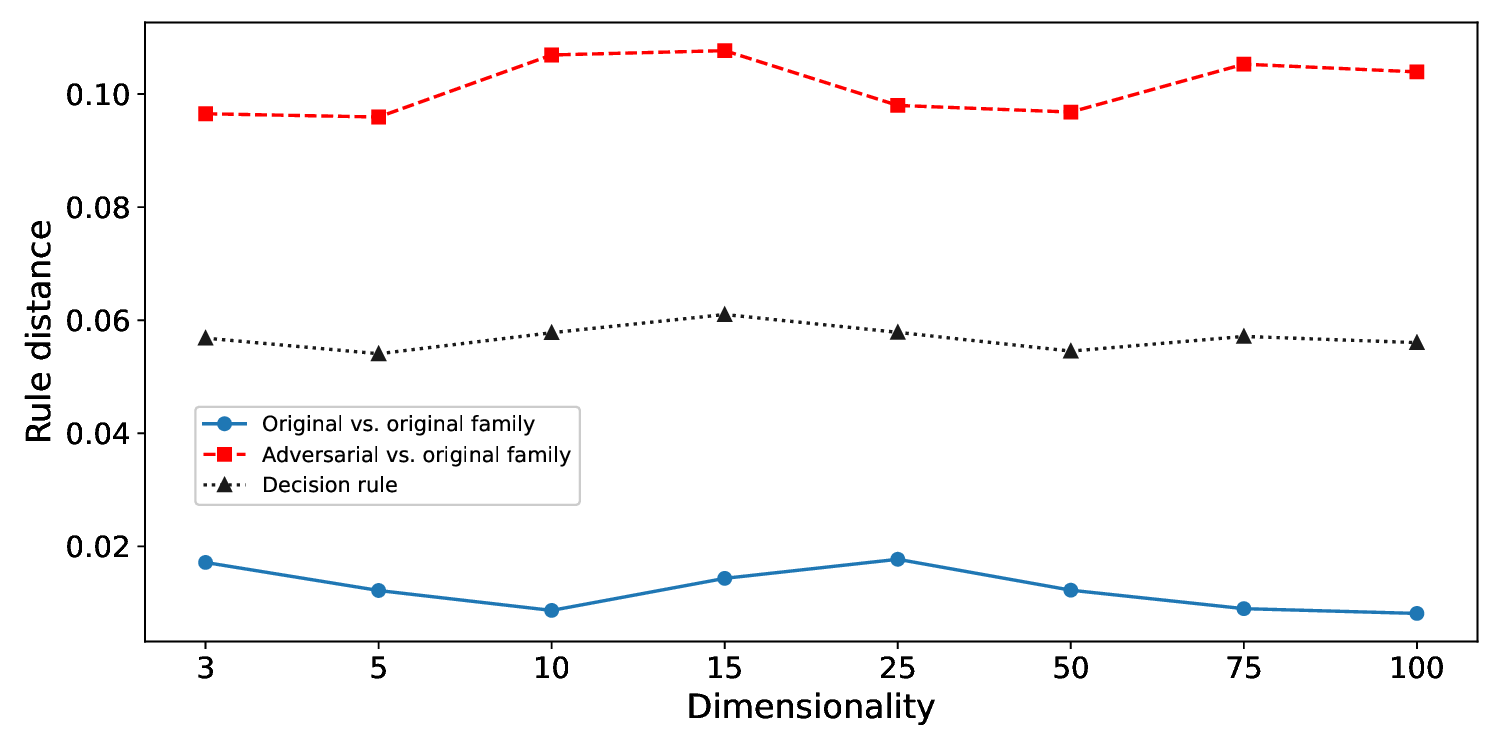}
        \caption{Agensla malware family}
    \end{subfigure}
    \hfill
    \begin{subfigure}[b]{0.47\textwidth}
        \centering
        \includegraphics[width=\textwidth]{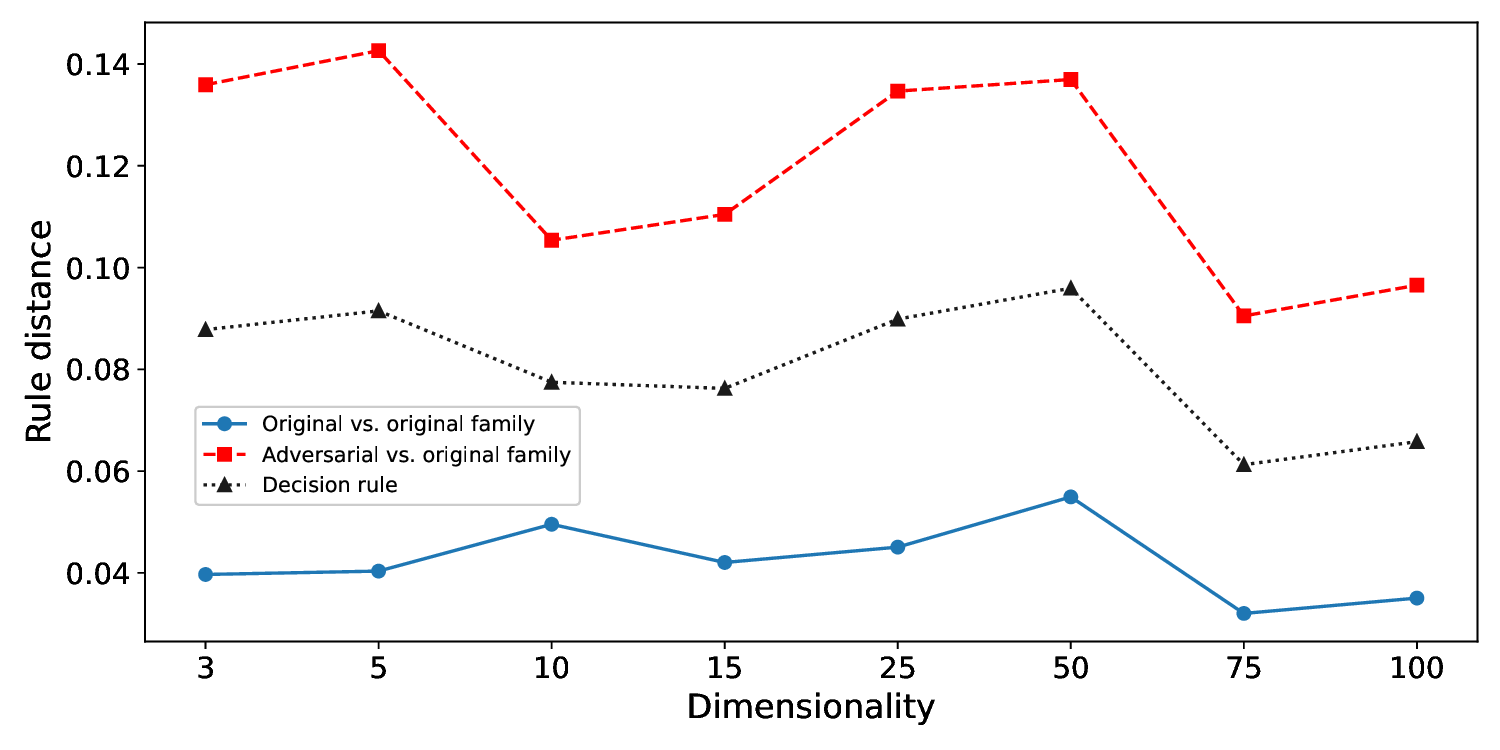}
        \caption{DCRat malware family}
    \end{subfigure}
    
    \vspace{0.5cm} % vertical space between rows

    % Second row
    \begin{subfigure}[b]{0.47\textwidth}
        \centering
        \includegraphics[width=\textwidth]{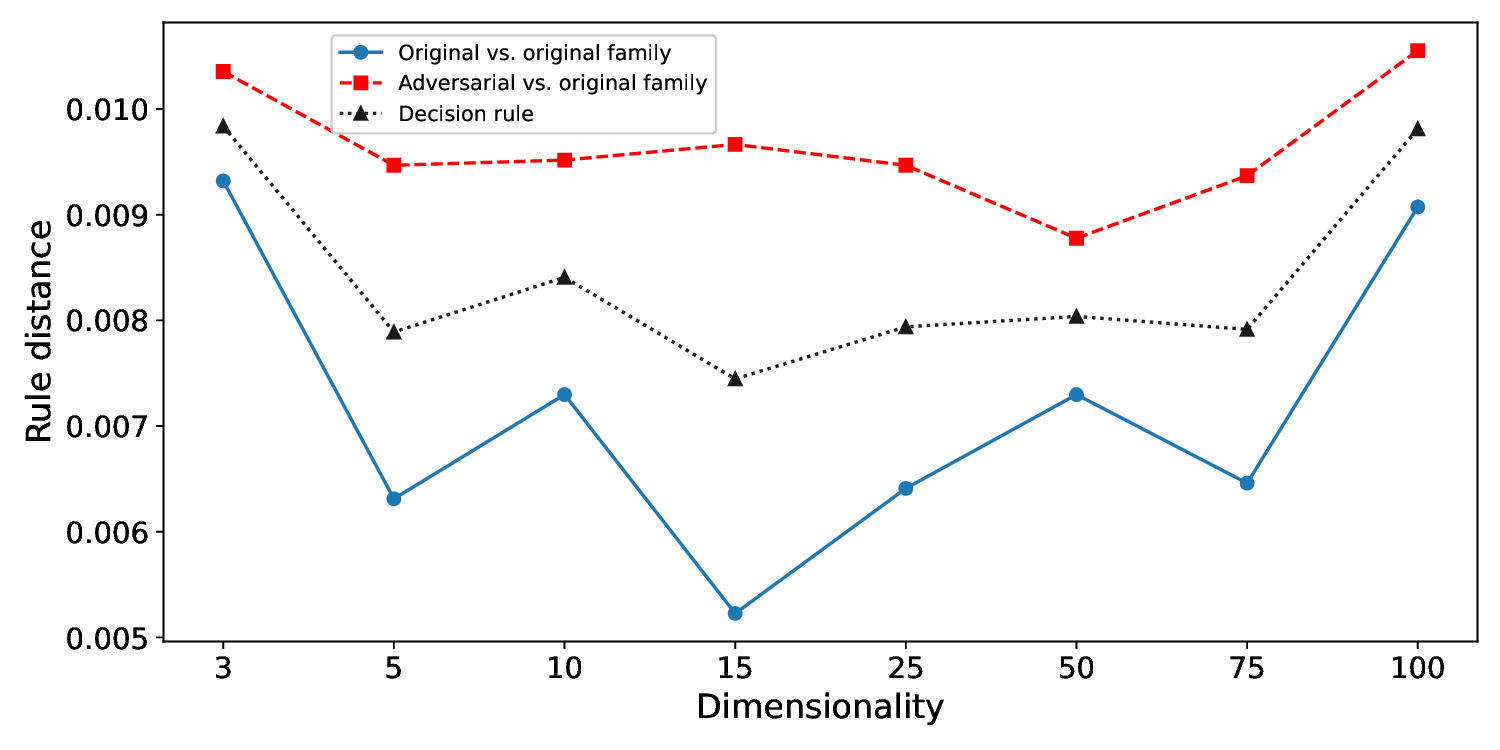}
        \caption{Makoob malware family}
    \end{subfigure}
    \hfill
    \begin{subfigure}[b]{0.47\textwidth}
        \centering
        \includegraphics[width=\textwidth]{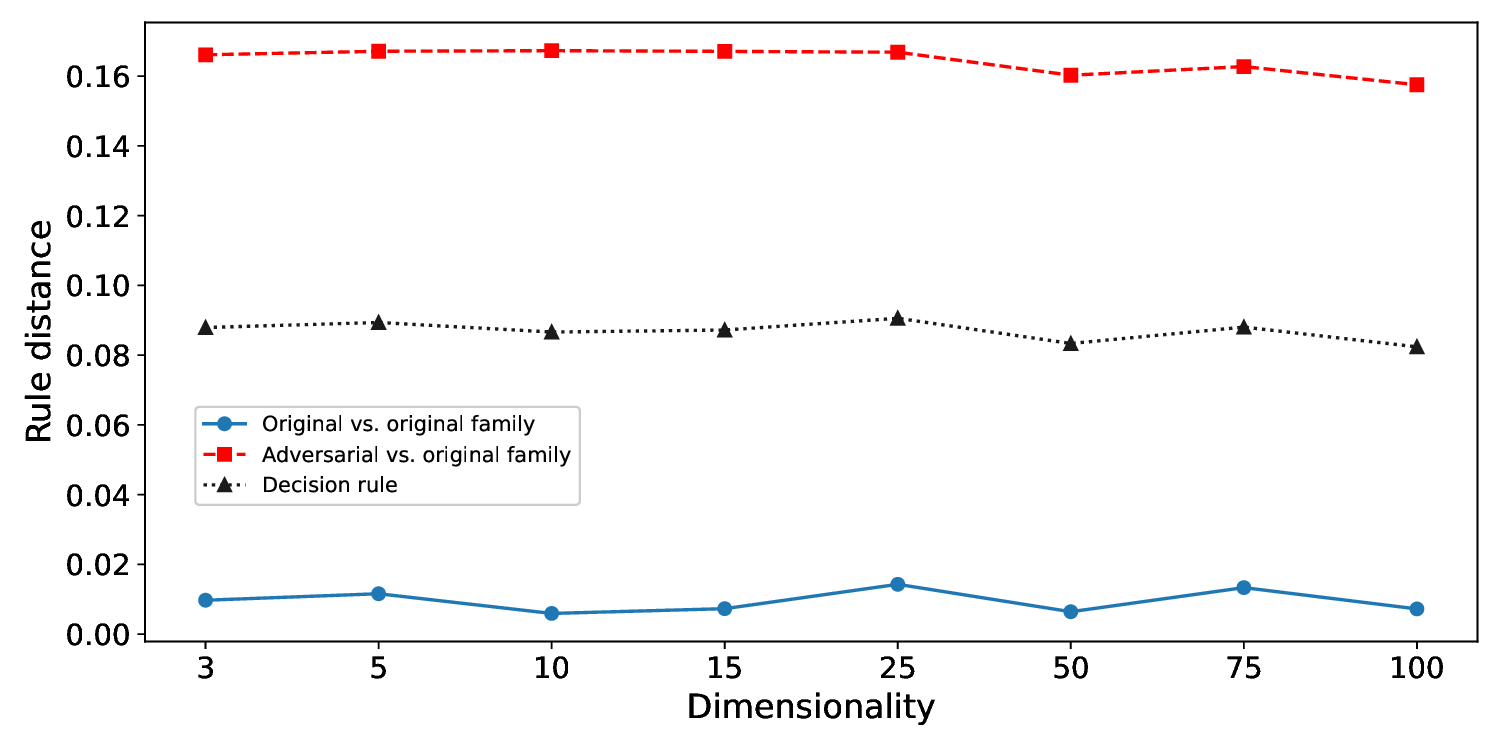}
        \caption{Mokes malware family}
    \end{subfigure}

    \vspace{0.5cm} % vertical space between rows

    % Third row
    \begin{subfigure}[b]{0.47\textwidth}
        \centering
        \includegraphics[width=\textwidth]{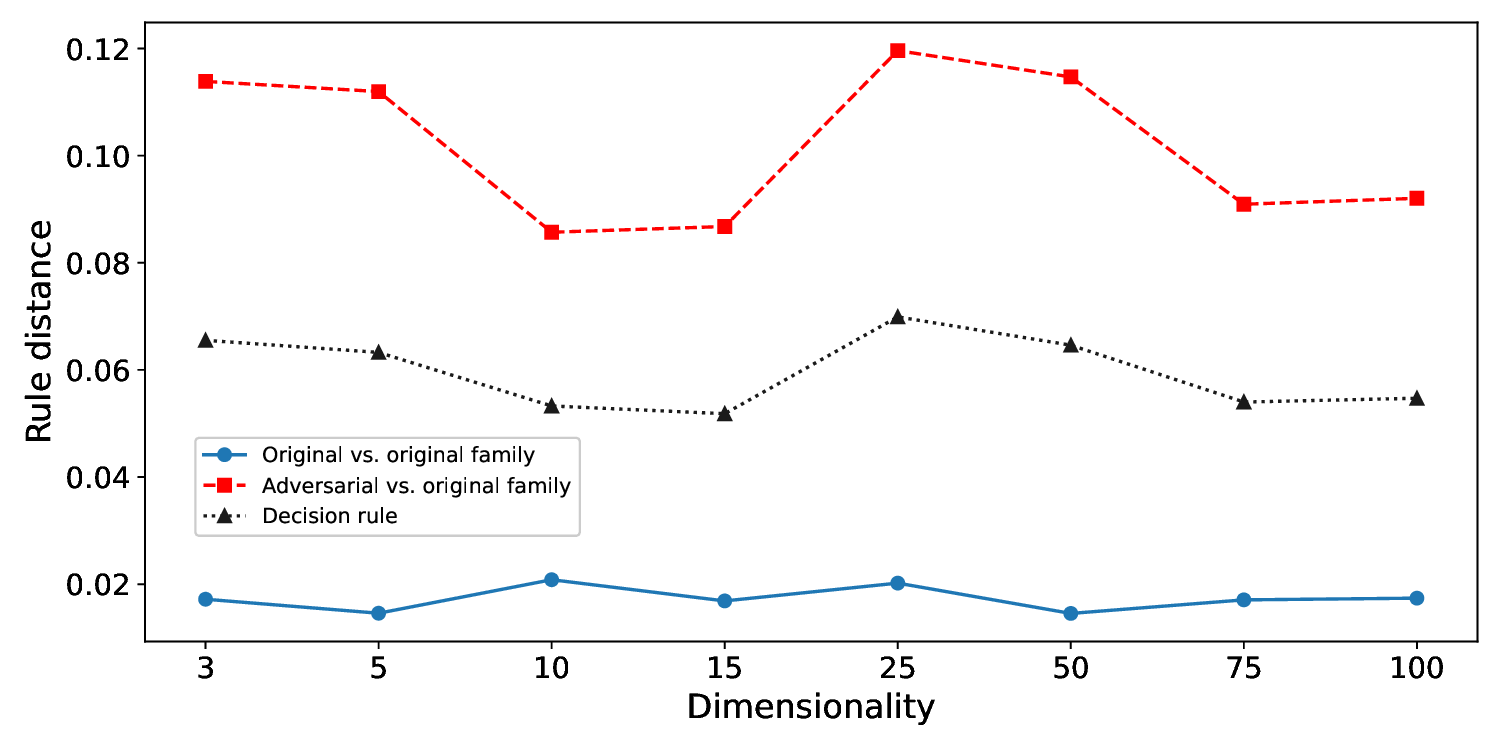}
        \caption{Strab malware family}
    \end{subfigure}
    \hfill
    \begin{subfigure}[b]{0.47\textwidth}
        \centering
        \includegraphics[width=\textwidth]{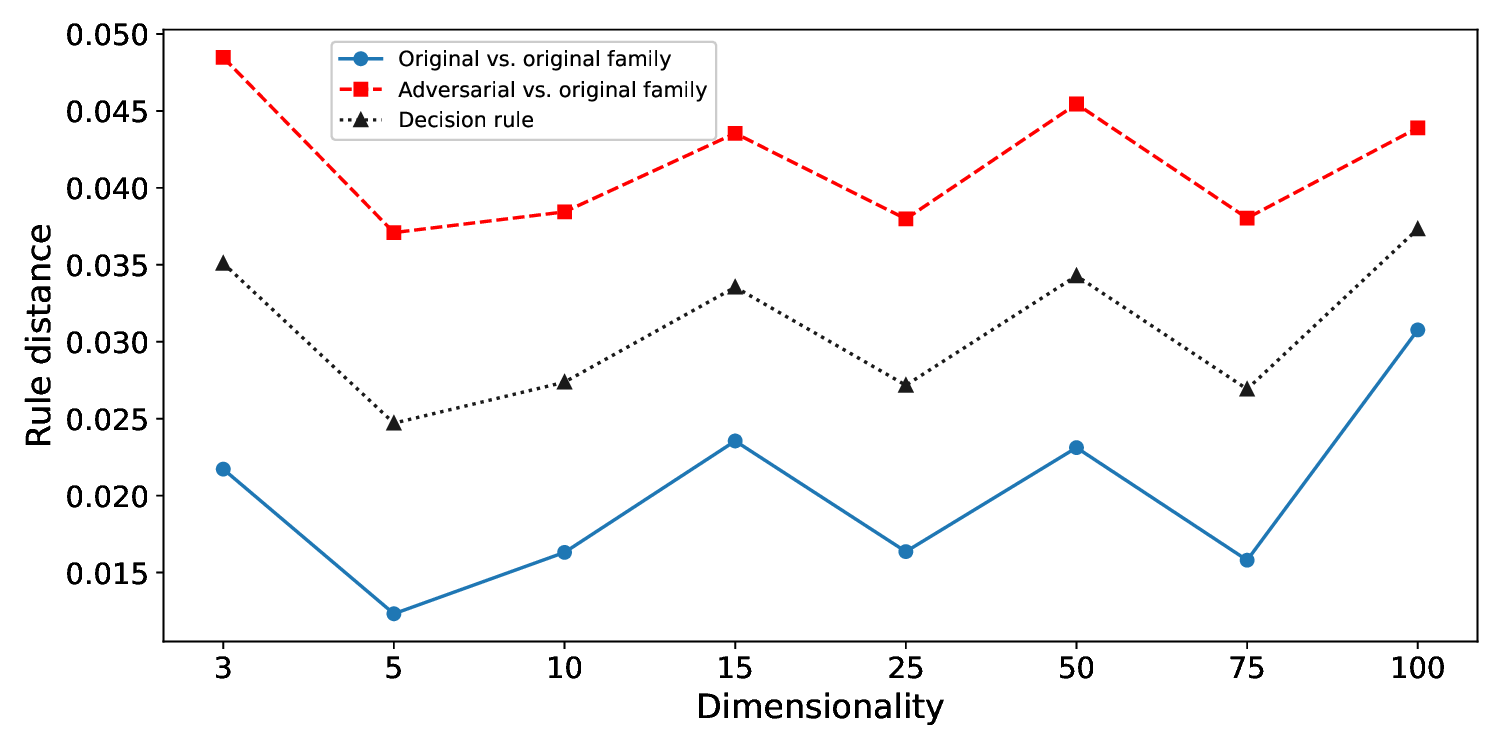}
        \caption{Taskun malware family}
    \end{subfigure}

    \caption{The distances between rules that are calculated only within the original family (Original vs. original family), and the distances between rules that are calculated from the original family and the rules computed from the adversarial family (Adversarial vs. original family). The Decision rule corresponds to the arithmetic mean of these two distance values and serves as a threshold for determining the presence of concept drift.}
    \label{fig:matrix3x2}
\end{figure*}

The procedure described above was carried out for six malware families and for the following feature vector dimensionalities: 3, 5, 10, 15, 25, 50, 75, and 100. Figure~\ref{fig:matrix3x2} illustrates a significant difference between two types of distances: the distances computed within the original family (i.e., comparing rule sets derived from different subsets of the same original family) and the distances computed between the original and adversarial families (i.e., comparing rule sets from the original family with those from its adversarially modified counterpart). For each family and each dimensionality, we performed 10 experiments, and the graphs report the mean. For the Mokes family, the largest difference between these distances is observed, indicating that concept drift is most effectively detected. Specifically, for Mokes, the distance between rule sets within the original family (i.e., comparing rule sets computed from \textit{set1} and \textit{set2}) is up to 15 times smaller than the distance between the original and adversarial families across all dimensionalities. On the other hand, for the Makoob family, the difference between rule distances was the smallest among the tested families.

Regarding Step~7, the decision on concept drift detection is based on the \textit{decision rule}, which is shown in Fig.~\ref{fig:matrix3x2}. The decision rule is defined as the arithmetic mean of the average distances for within-family comparisons (denoted as "Original vs. original family" in the figure) and the average distances for cross-family comparisons (denoted as "Adversarial vs. original family" in the figure). The procedure for detecting concept drift is as follows: 
\begin{enumerate}
    \item Apply the rule-based classifier RIPPER to the test set to generate new rules.
    \item Compute the distance between the newly generated rules and the most recent previously computed rules.
    \item If this distance exceeds the threshold defined by the \textit{decision rule}, predict that a concept drift has occurred.
\end{enumerate}

The procedure described for detecting concept drift assumes that we know the family to which each malware sample in the test set belongs. This means that a classification algorithm has already been applied to these samples, assigning them to a given malware family. In situations where the malware family is unknown, the approach proposed in \cite{jurevckova2024classification} can be employed, where a new malware family classification and clustering system is introduced, designed for the online processing of zero-day malware. Each sample is processed in real time and assigned either to an existing family or to a newly emerging malware family. For samples assigned to a newly emerging or previously unknown malware family, concept drift is not detected.

Table~\ref{tab:concept_drift_errors} shows the number of cases, out of a total of 10 experiments, in which concept drift was incorrectly detected using the proposed method. For the Agensla and Mokes families, concept drift was detected with 100\% accuracy across all feature vector dimensionalities. For the Taskun family, concept drift was incorrectly detected in only one out of 80 experiments, corresponding to a feature vector dimensionality of 100. The highest error rate in concept drift detection was observed for the DCRat family, amounting to 22.5\% when considering all feature vector dimensionalities; however, for a feature vector dimensionality of 25, the error rate was 0\%. The overall accuracy of concept drift detection across all six malware families and all feature vector dimensionalities was 92.08\%.

\begin{table}[htbp]
\centering
\caption{Number of errors in concept drift detection over ten experiments for different feature vector dimensionalities.}
\label{tab:concept_drift_errors}
\begin{tabular}{l|c|c|c|c|c|c|c|c}
\toprule
Family & 3 & 5 & 10 & 15 & 25 & 50 & 75 & 100 \\
\midrule
Agensla & 0 & 0 & 0 & 0 & 0 & 0 & 0 & 0 \\
DCRat   & 2 & 2 & 3 & 2 & 0 & 3 & 3 & 3 \\
Makoob  & 3 & 1 & 1 & 0 & 1 & 3 & 2 & 2 \\
Mokes   & 0 & 0 & 0 & 0 & 0 & 0 & 0 & 0 \\
Strab   & 0 & 2 & 1 & 0 & 1 & 1 & 1 & 0 \\
Taskun  & 0 & 0 & 0 & 0 & 0 & 0 & 0 & 1 \\
\bottomrule
\end{tabular}
\end{table}

The approach we propose not only allows for the detection of concept drift but also enables its quantification. The magnitude of the concept drift can be determined based on the distance between rules; that is, the greater the distance between the rules obtained from the most recent evolution of a family and the rules from the previous evolution, the more pronounced the concept drift.

The interpretability of concept drift detection lies in the use of rules, between which we compute distances to decide whether concept drift has occurred. We use a feature set that includes PE-format metadata, such as header information, section statistics, imported and exported functions, as well as byte-level distributions. These features then appear in the rules (see Section~\ref{pravidla} for an example), which clearly indicate which features and their values describe a given family. By comparing the rules from a newly evolved family with those from the previous iteration, we can identify which features and values are important for detecting the newly emerged family.

\section{\uppercase{Conclusions}}
\label{sec:conclusion}

Our findings demonstrate that rule-based drift analysis can effectively detect and explain evolutionary changes across malware families, achieving an overall concept drift detection accuracy of 92.08\% across six malware families. By comparing the generated rule sets, malware analysts can observe how the rules describing each family have changed over time, providing a clearer understanding of the evolution of a given malware family. As future work, we plan to extend this framework using decision trees, which offer an additional interpretable, rule-based mechanism for capturing and explaining malware behavior. Furthermore, experimenting with multiple types of distance metrics between rules could help enhance the interpretability of malware family evolution in terms of explicit changes in the rules themselves.

\section*{\uppercase{Acknowledgements}}

This work was supported by the Grant Agency of the CTU in Prague, grant No. SGS23/211/OHK3/3T/18 funded by the MEYS of the Czech Republic.

\bibliographystyle{apalike}
{\small
\bibliography{Example}}

\end{document}